\documentclass{iau} 
\usepackage{graphicx}

\title[S316.~~Self-shielding clumps] 
{Self-shielding clumps in starburst clusters}

\author[Jan Palou\v s, Richard W\" unsch, So\v na Ehlerov\' a \& Guillermo Tenorio-Tagle ]   
{Jan Palou\v s$^1$,
Richard W\" unsch$^1$,
So\v na Ehlerov\' a$^1$ \\ 
 \and Guillermo Tenorio-Tagle$^2$}

\affiliation{$^1$Astronomical Institute of the CAS, \\ Bo\v cn\' \i \ II 1401, 141 31 Prague 4, Czech Republic
\\ email: {\tt palous@ig.cas.cz} \\[\affilskip]
$^2$Instituto Nacional de Astrof\' \i sica Optica y Electronica, \\ AP 51, 72000 Puebla, Mexico 
\\email: {\tt guillermotenoriotagle@gmail.com}}

\pubyear{2015}
\volume{316}  
\jname{Formation, Evolution, and Survival of Massive Star Clusters}
\editors{C. Charbonnel \& A. Nota, eds.}
\begin{document}

\maketitle

\begin{abstract}
Young and massive star clusters above a critical mass form thermally unstable 
clumps reducing locally the temperature and pressure of the hot 10$^{7}$~K 
cluster wind. The matter reinserted by stars, and mass loaded in interactions 
with pristine gas and from evaporating circumstellar disks, accumulate on 
clumps that are ionized with photons produced by massive stars. We discuss
if they may become self-shielded when they reach the central part of the 
cluster, or even before it, during their free fall to the cluster center. Here we 
explore the importance of heating efficiency of stellar winds. 
\end{abstract}

\firstsection 
\section{Introduction}

High-precision photometry with Hubble Space Telescope and stellar spectroscopy with Very Large 
Telescope discovered multiple stellar generations in globular clusters: enhanced He content in blue 
main-sequences and anticorrelation of Na - O and other products of hydrogen burning in cluster stars, see 
\cite[Hempel \etal\ (2014)]{Hempel_etal14}, \cite[Gratton \etal (2012)]{Gratton_etal12} 
and \cite[Poitto \etal\ (2012)]{Poitto_etal12}. It is assumed that the first stellar generation mixed its 
nuclear burning products with original gas preparing the medium for formation of further stellar generations.
Possible explanations include slow winds of fast rotating massive stars (\cite[Krause \etal \ 2013]{Krause_etal13}), 
or winds of AGB stars (\cite[D'Ercole \etal \ 2012]{D'Ercole_etal12}). We propose the cooling winds of 
massive stars as the place where the enriched stellar generation forms. Here we discuss when thermally unstable clumps
self-shield the UV photons produced by young stars of the cluster to become seeds for secondary star formation. 
        
\section{Cooling Winds}

\begin{figure}[t]
 \vspace*{-0.3 cm}
\begin{center}
 \includegraphics[width=1.9in]{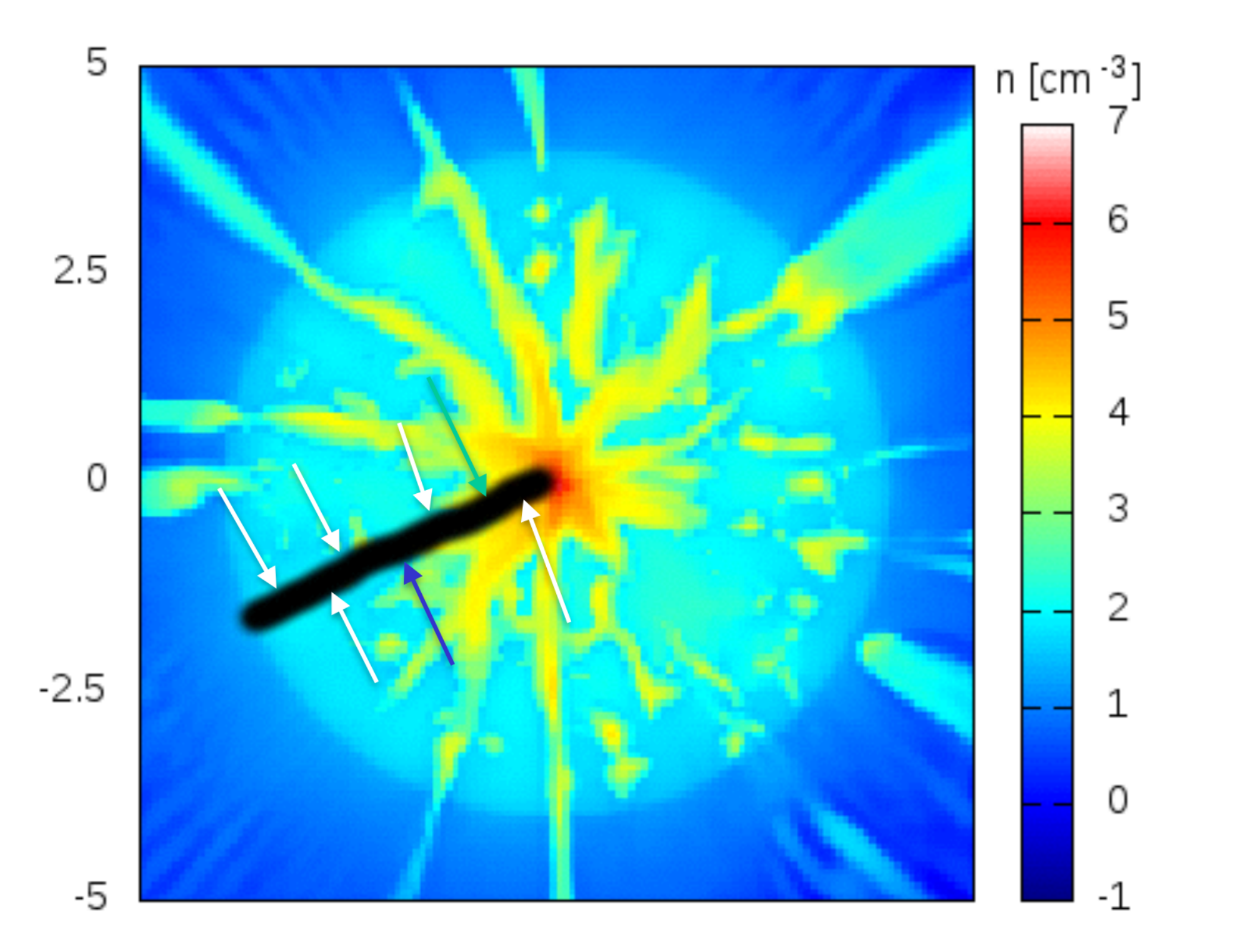}
 \includegraphics[width=1.9in]{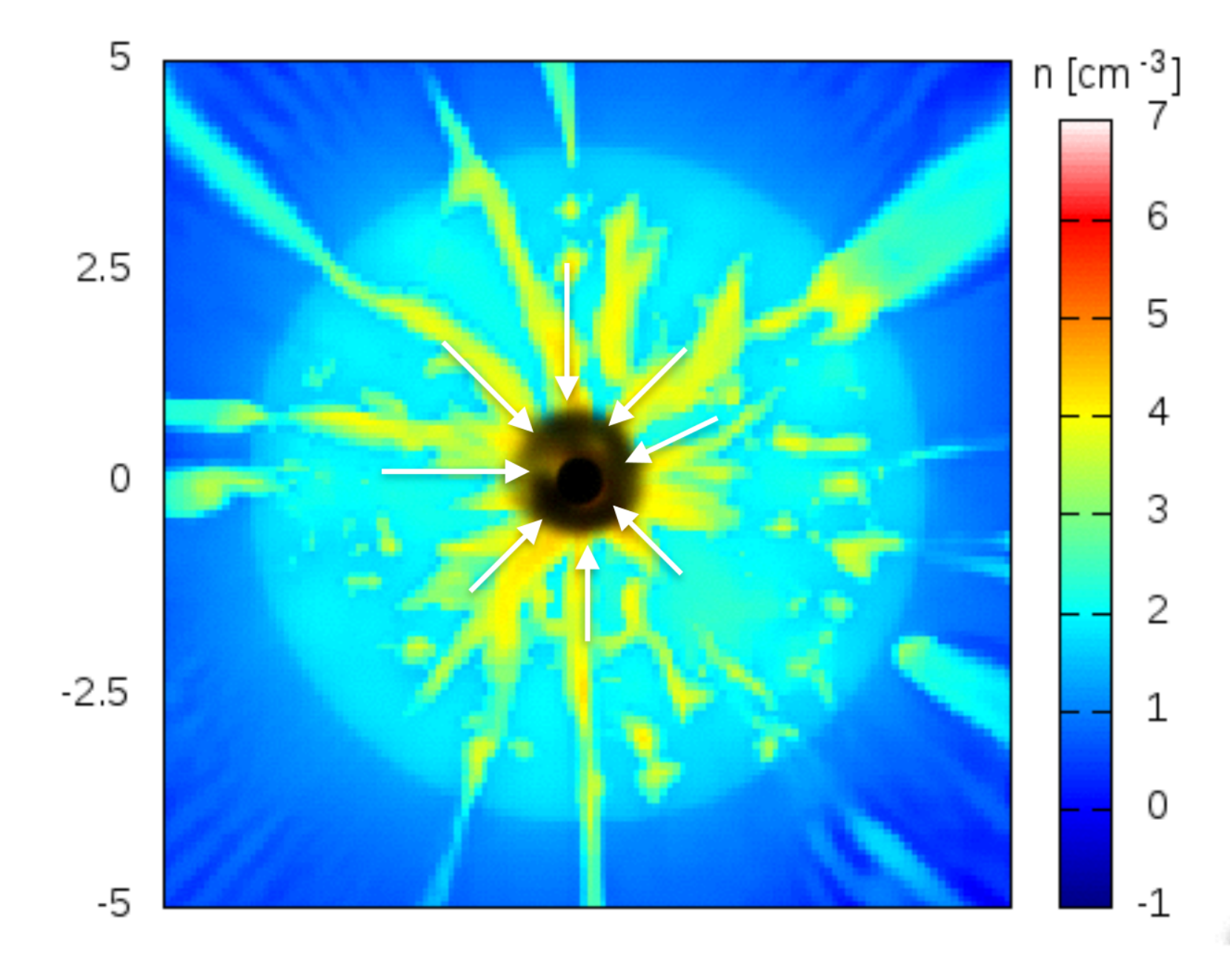}
 
 \vspace*{-0.3 cm}
 \caption{Left: stream of warm gas falling to the cluster center; Right: central condensation of warm gas.}
   \label{fig1}
\end{center}
\end{figure}

In clusters above a critical mass, thermally unstable clumps are created out of hot gas of fast winds of massive stars
(\cite[Tenorio-Tagle \etal\ 2007]{Tenorio_etal07}, \cite[W\"unsch \etal\  2011]{Wunsch_etal11}, 
\cite[Palou\v s \etal\  2013]{Palous)_etal13}, and references therein).
They shrink to a smaller volume, and the surrounding hot cluster wind is not able to push them out from the cluster, 
on the contrary, under the influence of the cluster gravity, they stream down towards the cluster center 
(see Fig. \ref{fig1}). We derive the minimum time that is needed to accumulate enough mass so that the stream is 
self-shielded against UV photons from the star cluster. However, when the free fall time of clumps is shorter compared 
to the self-shielding time, the streams of thermally unstable clumps meet in the cluster center forming central 
concentration of warm gas. As soon as this central mass concentration is self-shielded, it cools further down to become 
the seed of a new enriched stellar generation.

\vspace*{-0.5cm}
\section{Results}
\begin{figure}
\begin{center}
 \includegraphics[width=2.6in]{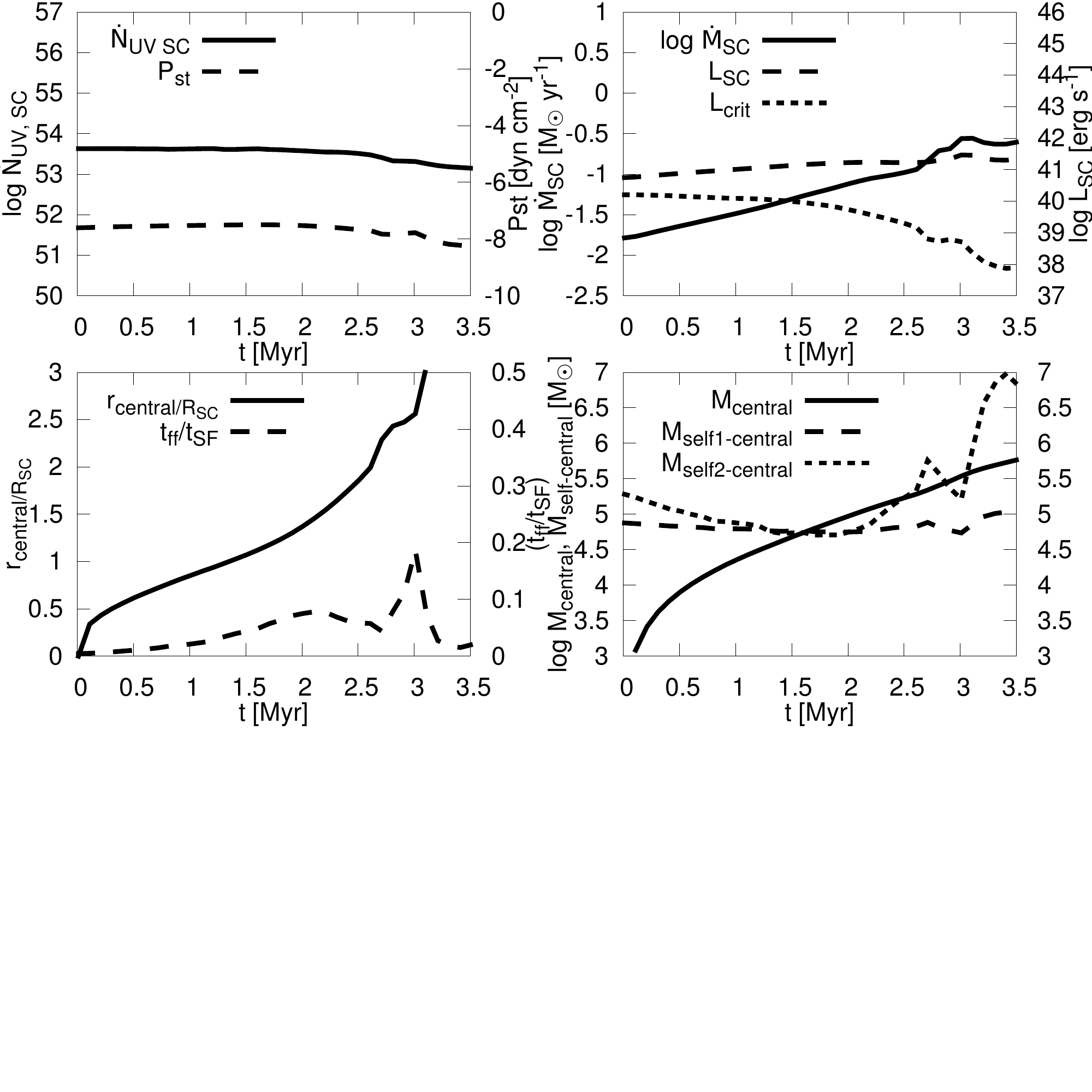}
 \includegraphics[width=2.6in]{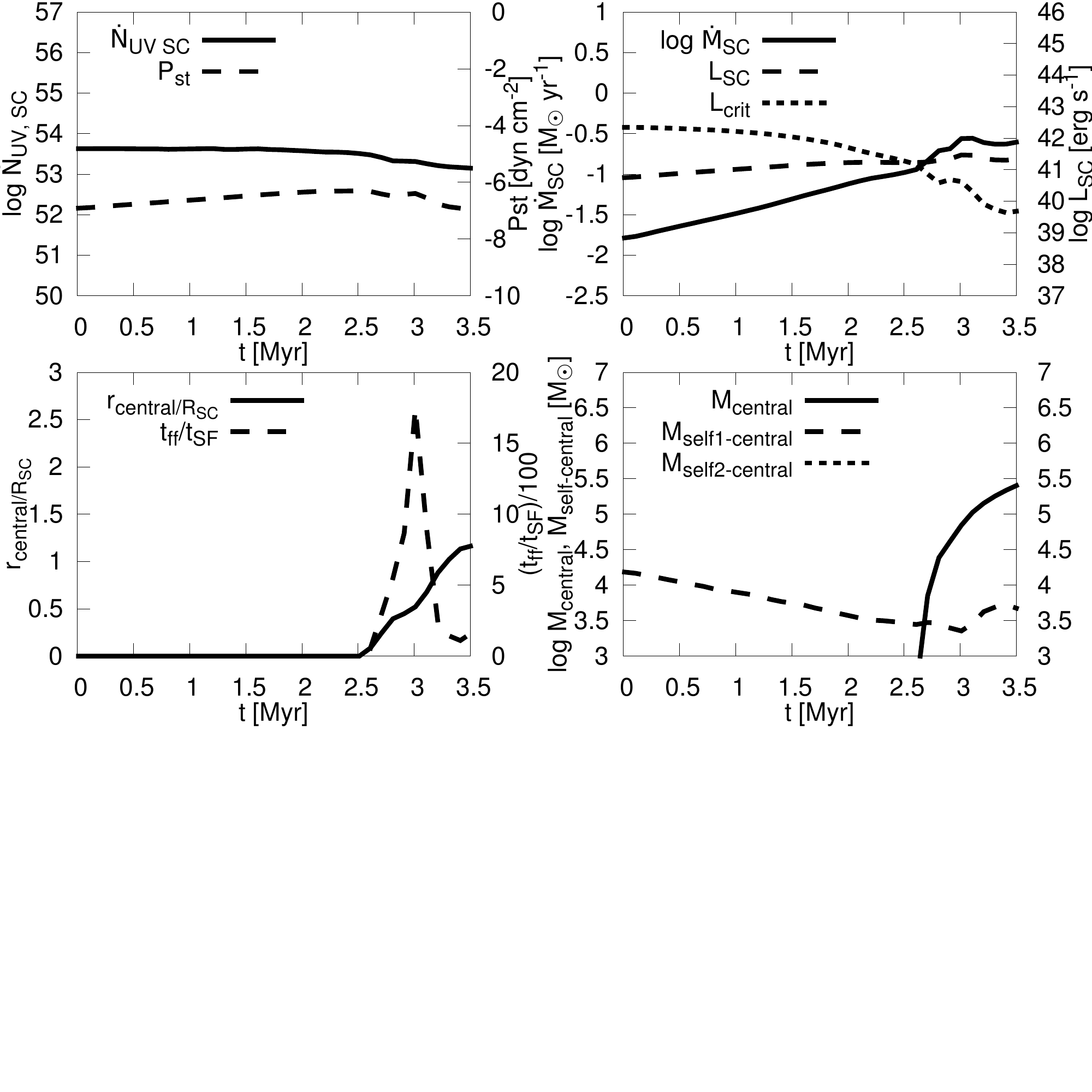}
 \vspace*{-2.4 cm}
 \caption{The time evolution of the UV photon flux $\dot N_{UV, SWC}$, pressure in the cluster wind $P_{st}$, 
total mass flux $\dot M_{SC}$ and mechanical 
energy flux $L_{SC}$ in the cluster wind, radius of the central clump $r_{central}$, 
self-shielding time of the gas streams $t_{SF}$ and 
central mass concentration $M_{central}$ during 
the first 3.5 Myr. Left: 5\% heating efficiency; Right 30\% heating efficiency.}
   \label{fig2}
\end{center}
\end{figure}

The results of our semi-analytical calculations of the evolution of clusters during its first 3.5 Myr 
are given in Fig. \ref{fig2}. We assume a first stellar generation with the total 
mass $10^7$ M$_\odot$ and a radius 3 pc. In such cluster, its wind is thermally unstable forming after 3.5 Myr 
clumps with a total mass of a few times 10$^5$ M$_\odot$. An important parameter of the model is the heating efficiency, 
giving the fraction of the wind mechanical energy that is converted into heat of the hot cluster winds. 
In the case of a low heating efficiency ($\sim$5 \%) the free-fall time is shorter compared to the self-shielding time, 
forming centrally concentrated second stellar generation. On the other hand, when the heating efficiency is high 
($\sim$30\%) the gas streams become self shielded before they reach the cluster center, resulting in a  second generation 
cluster with a radius similar to that of the first stellar generation.  

\noindent
{\it Acknowledgements.} This study has been supported by the Czech Science 
Foundation grant 209/15/06012S and by the project RVO: 6785815.

\end{document}